\documentclass{article}
\usepackage{times}
\begin{document}
\title{Quantum States from Tangent Vectors}
\author{Jos\'e M. Isidro\\
Instituto de F\'{\i}sica Corpuscular (CSIC--UVEG)\\
Apartado de Correos 22085, Valencia 46071, Spain\\
and\\
Max-Planck-Institut f\"ur Gravitationsphysik, 
\\Albert-Einstein-Institut, \\
D-14476 Golm, Germany\\
{\tt jmisidro@ific.uv.es}}

\maketitle

\begin{abstract}
We argue that tangent vectors to classical phase space give rise to quantum states
of the corresponding quantum mechanics. This is established for the case of
complex, finite--dimensional, compact, classical phase spaces ${\cal C}$, 
by explicitly constructing Hilbert--space vector bundles over 
${\cal C}$. We find that these vector bundles split as the direct sum of two holomorphic 
vector bundles: the holomorphic tangent bundle $T({\cal C})$, plus a complex line bundle 
$N({\cal C})$. Quantum states (except the vacuum) appear as tangent vectors to ${\cal C}$. 
The vacuum state appears as the fibrewise generator of $N({\cal C})$. 
Holomorphic line bundles $N({\cal C})$ are classified by the elements of ${\rm Pic}\,({\cal C})$, 
the Picard group of ${\cal C}$. In this way ${\rm Pic}\,({\cal C})$ appears as the parameter space 
for nonequivalent vacua. Our analysis is modelled on, but not limited to, the case when ${\cal C}$ 
is complex projective space ${\bf CP}^n$. 

2001 Pacs codes: 03.65.Bz, 03.65.Ca, 03.65.-w.

\end{abstract}

\tableofcontents

\section{Introduction}\label{intro}

{}Fibre bundles are powerful tools to formulate the gauge theories 
of fundamental interactions and gravity \cite{LIBAZCA}. The question arises whether or not 
quantum mechanics may also be formulated using fibre bundles. Important physical motivations call for such a formulation.

In quantum mechanics one aims at contructing a Hilbert--space vector bundle 
over classical phase space. In geometric quantisation  
this goal is achieved in a two--step process that can be very succintly summarised 
as follows. One first constructs a certain holomorphic line bundle (the {\it quantum line
bundle}\/) over classical phase space. Next one identifies certain sections of this line bundle  
as defining the Hilbert space of quantum states. Alternatively one may skip 
the quantum line bundle and consider the one--step process of directly constructing a 
Hilbert--space vector bundle over classical phase space. Associated with this vector bundle 
there is a principal bundle whose fibre is the unitary group of Hilbert space.

Standard presentations of quantum mechanics usually deal with the case when this 
Hilbert--space vector bundle is trivial. Such is the case, 
{\it e.g.}, when classical phase space is contractible to a point. However, 
it seems natural to consider the case of a nontrivial bundle as well. 
Beyond a purely mathematical interest, important physical issues that go 
by the generic name of {\it dualities} \cite{VAFA} motivate 
the study of nontrivial bundles. 

Triviality of the Hilbert--space vector bundle implies that the transition 
functions all equal the identity of the structure group. In passing from 
one coordinate chart to another on classical phase space, vectors on the 
fibre are acted on by the identity. Since these vectors are quantum 
states, we can say that all observers on classical phase space are quantised 
in the same way. This is no longer the case on a nontrivial vector bundle, 
where the transition functions are different from the identity.
As opposed to the previous case, different neighbourhoods 
on classical phase space are quantised independently and, possibly, differently. 
The resulting quantisation is only local on classical phase space, instead of global.
This reflects the property of local triviality satisfied by all fibre bundles.

Given a certain base manifold and a certain fibre, the trivial bundle over the given base 
with the given fibre is unique. This may mislead one to conclude that quantisation 
is also unique, or independent of the observer on classical phase space. In 
fact the notion of duality points precisely to the opposite conclusion, {\it i.e.}, 
to the nonuniqueness of the quantisation procedure and to its dependence on the observer 
\cite{VAFA}. 

Clearly a framework is required in order to accommodate dualities within quantum mechanics 
\cite{VAFA}. Nontrivial Hilbert--space vector bundles over classical phase space provide 
one such framework. They allow for the possibility of having different, nonequivalent quantisations,
as opposed to the uniqueness of the trivial bundle. However, although nontriviality 
is a necessary condition, it is by no means sufficient. A flat connection on a nontrivial 
bundle would still allow, by parallel transport, to canonically identify the Hilbert--space 
fibres above different points on classical phase space. This identification would depend 
only on the homotopy class of the curve joining the basepoints, but not on the curve itself. 
Now flat connections are characterised by {\it constant}\/ transition functions \cite{KN}, 
this constant being always the identity in the case of the trivial bundle. 
Hence, in order to accommodate dualities, we will be looking for {\it nonflat}\/ connections. 
We will see presently what connections we need on these bundles.

This article is devoted to constructing nonflat Hilbert--space vector bundles over classical 
phase space. In motivating the subject we have dealt with unitary groups as structure groups 
and linear fibres such as Hilbert spaces. However quantum states are rays rather than vectors. 
Therefore it is more precise to consider the corresponding {\it projective}\/ spaces and 
{\it projective}\/ unitary groups, as we will do from now on.

Throughout this article, ${\cal C}$ will denote a complex $n$--dimensional, 
connected, compact classical phase space, endowed with a symplectic form $\omega$ 
and a complex structure ${\cal J}$. We will assume that $\omega$ and ${\cal J}$ 
are compatible, so holomorphic coordinate charts on ${\cal C}$ will also 
be Darboux charts. We will mostly concentrate on the case when ${\cal C}$ 
is projective space ${\bf CP}^n$. Its holomorphic tangent bundle will be denoted 
$T({\bf CP}^n)$. The following line bundles over ${\bf CP}^n$ will be considered:
the trivial line bundle $\epsilon$, the tautological line bundle $\tau^{-1}$ 
and its dual $\tau$. The Picard group of ${\cal C}$ will be denoted 
${\rm Pic}\,({\cal C})$. ${\cal H}$ will denote the complex, $(N+1)$--dimensional 
Hilbert space of quantum states ${\bf C}^{N+1}$, with unitary group $U(N+1)$. 
They projectivise to ${\bf CP}^{N}$ and $PU(N)$, respectively. 

Our analysis will deal mostly with the case when ${\cal C}={\bf CP}^n$. 
In section \ref{cipienne} we summarise its useful properties as a classical 
phase space. In section \ref{cipienne} we recall some well--known facts from 
geometric quantisation. They concern the dimension of the space of holomorphic 
sections of the quantum line bundle on a compact, quantisable K\"ahler 
manifold. This dimension is rederived in section \ref{esstqmb} using purely 
quantum--mechanical arguments, by constructing the Hilbert--space bundle 
of quantum states over ${\bf CP}^n$. For brevity, the following summary 
deals only with the case when the Hilbert space is ${\bf C}^{n+1}$
(see sections \ref{wrepp}, \ref{xcomptx} for the general case).
The fibre ${\bf C}^{n+1}$ over a given coordinate chart 
on ${\bf CP}^n$ is spanned by the vacuum state $|0\rangle$, 
plus $n$ states $A^{\dagger}_j|0\rangle$,
$j=1,\ldots, n$, obtained by the action of creation operators. We identify 
the transition functions of this bundle as jacobian matrices plus a phase factor. 
The jacobian matrices account for the transformation (under coordinate changes on 
${\bf CP}^n$) of the states $A^{\dagger}_j|0\rangle$, while the phase factor 
corresponds to $|0\rangle$. This means that all quantum states (except the vacuum) 
are tangent vectors to ${\bf CP}^n$. In this way the Hilbert--space bundle over 
${\bf CP}^n$ splits as the direct sum of two holomorphic vector bundles: 
the tangent bundle $T({\bf CP}^n)$, plus a line bundle $N({\bf CP}^n)$ 
whose fibrewise generator is the vacuum. 

All complex manifolds admit a Hermitian metric, so having tangent vectors 
as quantum states suggests using the Hermitian connection and the corresponding 
curvature tensor to measure flatness. Now $T({\bf CP}^n)$ is nonflat,
so it fits our purposes. The freedom in having different nonflat Hilbert--space bundles 
over ${\bf CP}^n$ resides 
in the different possible choices for the complex line bundle $N({\bf CP}^n)$. Such choices 
are 1--to--1 with the elements of the Picard group ${\rm Pic}\,({\bf CP}^n)={\bf Z}$.  

The previous picture of quantum states (except the vacuum) as tangent vectors
remains substantially correct in the case of an arbitrary, compact, complex manifold 
${\cal C}$ whose complex and symplectic structures are compatible;
this is proved in section \ref{ttvvqqss}.
Flatness of the resulting Hilbert--space bundle depends on whether or not the 
holomorphic tangent bundle $T({\cal C})$ is flat. We continue to have the Picard group 
${\rm Pic}\,({\cal C})$ as the parameter space for different Hilbert--space bundles 
over ${\cal C}$. Finally section \ref{dicu} discusses our results. 

Topics partially overlapping with ours are dealt with in refs. \cite{MATONE, MINIC, CARROLL, MEX, MPLA}.

\section{Properties of ${\bf CP}^n$ as a classical phase space}\label{cipienne}

We will consider a classical mechanics whose phase space ${\cal C}$ is 
complex, projective $n$--dimensional space ${\bf CP}^n$.
The following properties are well known \cite{KN}.

Let $Z^1,\ldots, Z^{n+1}$ denote homogeneous coordinates on ${\bf CP}^n$. 
The chart defined by $Z^k\neq 0$ covers one copy of the open set 
${\cal U}_k={\bf C}^n$. On the latter we have the holomorphic coordinates 
$z^j_{(k)}=Z^j/Z^k$, $j\neq k$; there are $n+1$ such coordinate charts. 
${\bf CP}^n$ is a K\"ahler manifold with respect 
to the Fubini--Study metric. On the chart $({\cal U}_k, z_{(k)})$ the K\"ahler 
potential reads
\begin{equation}
K(z^j_{(k)}, {\bar z}^j_{(k)})=
\log{\left(1 + \sum_{j=1}^n z^j_{(k)} {\bar z}^j_{(k)}\right)}.
\label{fubst}
\end{equation} 
The singular homology ring $H_*\left({\bf CP}^n, {\bf Z}\right)$ contains 
the nonzero subgroups 
\begin{equation}
H_{2k}\left({\bf CP}^n, {\bf Z}\right)={\bf Z}, \qquad 
k=0,1,\ldots, n,
\label{oncero}
\end{equation}
while 
\begin{equation}
H_{2k+1}\left({\bf CP}^n, {\bf Z}\right)=0, \qquad 
k=0,1,\ldots, n-1.
\label{oncerox}
\end{equation}
We have ${\bf CP}^{n}={\bf C}^n\cup {\bf CP}^{n-1}$, with ${\bf CP}^{n-1}$ a hyperplane 
at infinity. Topologically, ${\bf CP}^{n}$ is obtained by attaching a (real) $2n$--dimensional 
cell to ${\bf CP}^{n-1}$. ${\bf CP}^n$ is simply connected,
\begin{equation}
\pi_1\left({\bf CP}^n\right)=0,
\label{grfund}
\end{equation}
it is compact, and inherits its complex structure from that on ${\bf C}^{n+1}$. 
It can be regarded as the Grassmannian manifold 
\begin{equation}
{\bf CP}^n=U(n+1)/\left(U(n)\times U(1)\right)=S^{2n+1}/U(1).
\label{facx}
\end{equation}

Let $\tau^{-1}$ denote the {\it tautological bundle}\/ on ${\bf CP}^n$. We recall that 
$\tau^{-1}$ is defined as the subbundle of the trivial bundle ${\bf CP}^n\times 
{\bf C}^{n+1}$ whose fibre at $p\in {\bf CP}^n$ is the line in ${\bf C}^{n+1}$ 
represented by $p$. Then $\tau^{-1}$ is a holomorphic line bundle over ${\bf CP}^n$. 
Its dual, denoted $\tau$, is called the {\it hyperplane bundle}. 
For any $l\in {\bf Z}$, the $l$--th power $\tau ^l$ is also a holomorphic line bundle 
over ${\bf CP}^n$. In fact every holomorphic line bundle 
$L$ over ${\bf CP}^n$ is isomorphic to $\tau ^l$ for some $l\in {\bf Z}$;
this integer is the first Chern class of $L$. 

In the framework of geometric quantisation 
it is customary to consider the case when ${\cal C}$ is a compact 
K\"ahler manifold. In this context one introduces the notion of a quantisable, 
compact, K\"ahler phase space ${\cal C}$, of which ${\bf CP}^n$ is an 
example. This means that there exists a {\it quantum line bundle}\/ 
$({\cal L}, g, \nabla)$ on ${\cal C}$, where ${\cal L}$ is a holomorphic line bundle, 
$g$ a Hermitian metric on ${\cal L}$, and $\nabla$ a covariant derivative compatible 
with the complex structure and $g$. Furthermore, the curvature 
$F$ of $\nabla$ and the symplectic 2--form $\omega$ are required to satisfy
\begin{equation}
F=-2\pi {\rm i} \omega.
\label{quanxti}
\end{equation}
It turns out that quantisable, compact K\"ahler manifolds are projective 
algebraic manifolds and viceversa \cite{SCHLICHENMAIER}. After introducing 
a polarisation, the Hilbert space of quantum states is given by 
the global holomorphic sections of ${\cal L}$.

Recalling that, on ${\bf CP}^n$, ${\cal L}$ is isomorphic to $\tau^l$ for some 
$l\in {\bf Z}$, let ${\cal O}(l)$ denote the sheaf of holomorphic sections  
of ${\cal L}$ over ${\bf CP}^n$. The vector space of holomorphic sections of 
${\cal L}=\tau^l$ is the sheaf cohomology space $H^0({\bf CP}^n, {\cal O}(l))$.
The latter is zero for $l<0$, while for $l\geq 0$ it can be canonically 
identified with the set of homogeneous polynomials of degree $l$ on ${\bf C}^{n+1}$.
This set is a vector space of dimension $\left({n+l\atop 
n}\right)$:
\begin{equation}
{\rm dim}\,H^0({\bf CP}^n, {\cal O}(l))=\left({n+l\atop n}\right).
\label{jjoo}
\end{equation}
We will give a quantum--mechanical derivation of eqn. (\ref{jjoo}) in 
section \ref{esstqmb}.

Equivalence classes of holomorphic line bundles over a complex manifold 
${\cal C}$ are classified by the Picard group ${\rm Pic}\,({\cal C})$. 
The latter is defined \cite{LIBSCHL} as the sheaf cohomology group 
$H^1_{\rm sheaf}({\cal C}, {\cal O}^*)$,
where ${\cal O}^*$ is the sheaf of nonzero holomorphic functions on ${\cal C}$.
When ${\cal C}={\bf CP}^n$ things simplify because the above
sheaf cohomology group is in fact isomorphic to a singular homology group,
\begin{equation}
H^1_{\rm sheaf}({\bf CP}^n, {\cal O}^*)=H^2_{\rm sing}({\bf CP}^n, {\bf Z}),
\label{tsmplif}
\end{equation}
and the latter is given in eqn. (\ref{oncero}). Thus
\begin{equation}
{\rm Pic}\,({\bf CP}^n)={\bf Z}.
\label{athh}
\end{equation}
The zero class corresponds to the trivial line bundle $\epsilon=\tau^0$;
all other classes  correspond to nontrivial bundles. 
As the equivalence class of ${\cal L}$ varies, 
so does the space ${\cal H}$ of its holomorphic sections vary.

\section{Quantum Hilbert--space bundles over ${\bf CP}^n$}\label{esstqmb}

As discussed in section \ref{intro}, in quantum mechanics one skips the 
quantum line bundle ${\cal L}$ of geometric quantisation and proceeds directly to 
construct Hilbert--space bundles over classical phase space. We will therefore 
analyse such vector bundles (that we will call {\it quantum Hilbert--space bundles}, 
or ${\cal QH}$--bundles for short), their principal unitary bundles and, 
finally, their projectivisations. Our aim is to demonstrate that there are different 
nonequivalent choices for the nonflat ${\cal QH}$--bundles, to study how the corresponding 
quantum mechanics varies with each choice, and to provide a physical interpretation.
Although we will be able to reproduce the results that geometric quantisation derives 
from ${\cal L}$, our approach will be based on the ${\cal QH}$--bundles instead.
In particular, triviality of the quantum line bundle ${\cal L}$ does not imply, 
nor is implied by, triviality of the ${\cal QH}$--bundle; the same 
applies to flatness. 

Our analysis will be modelled on the case when ${\cal C}={\bf CP}^n$. 
An example of a classical dynamics on ${\bf CP}^n$ is given by the projective 
oscillator. On the coordinate chart ${\cal U}_k$ of eqn. (\ref{fubst}), 
the classical Hamiltonian equals the K\"ahler potential (\ref{fubst}).
Compactness of ${\bf CP}^n$ implies that, upon quantisation, the Hilbert space 
${\cal H}$ is finite--dimensional, and hence isomorphic 
to ${\bf C}^{N+1}$ for some $N$. This property follows from the fact that 
the number of quantum states grows monotonically with the symplectic volume 
of ${\cal C}$; the latter is finite when ${\cal C}$ is compact.
We are thus led to considering principal $U(N+1)$--bundles 
over ${\bf CP}^n$ and to their classification. Equivalently, we will consider 
the associated holomorphic vector bundles with fibre ${\bf C}^{N+1}$.
The corresponding projective bundles are ${\bf CP}^N$--bundles and principal $PU(N)$--bundles. 
Each choice of a different equivalence class of bundles will give rise to a different 
quantisation. How many such equivalence classes are there? For the moment let us observe that there is more than one. 
For example one can consider the class of the trivial bundle 
${\bf CP}^n\times U(N)$, or the class of a nontrivial bundle over ${\bf CP}^n$ 
such as the Hopf bundle. For the same reasons we can expect more than one equivalence 
class of projective bundles to exist. That this is actually true will also be proved 
later on.

So far we have left $N$ undetermined. In order to fix it we first 
pick the symplectic volume form $\omega^n$ on ${\bf CP}^n$ such that 
\begin{equation}
\int_{{\bf CP}^n}\omega^n=n+1.
\label{boludo}
\end{equation}
Next we set $N=n$, so ${\rm dim}\,{\cal H}=n+1$. This normalisation corresponds 
to 1 quantum state per unit of symplectic volume on ${\bf CP}^n$. Thus, {\it e.g.}, 
when $n=1$ we have the Riemann sphere ${\bf CP}^1$ and ${\cal H}={\bf C}^2$. 
The latter is the Hilbert space of a spin $s=1/2$ system, and the counting 
of states is correct. There are a number of further advantages to this normalisation. 
In fact eqn. (\ref{boludo}) is more than just a normalisation, in the sense that the 
dependence of the right--hand side on $n$ is determined by physical consistency arguments. 
This will be explained in section \ref{xcompt}.
Normalisation arguments can enter eqn. (\ref{boludo}) only through overall numerical 
factors such as $2\pi$, i$\hbar$, or similar. It is these latter factors that we fix by hand 
in eqn. (\ref{boludo}). 

The right--hand of our normalisation (\ref{boludo}) differs from 
that corresponding to eqn. (\ref{quanxti}). Up to numerical factors such as 
$2\pi$, ${\rm i}\hbar$, etc, it is standard  to set $\int_{{\bf CP}^n} F^n = n$ \cite{KN}.
However we will find our normalisation (\ref{boludo}) more convenient. 
Indeed we will make no use of the quantum line bundle ${\cal L}$, while we 
will be able to reproduce quantum--mechanically the results of geometric quantisation.

\subsection{Computation of ${\rm dim}\,H^0({\bf CP}^n, {\cal O}(1))$}\label{xcompt}

Next we present a quantum--mechanical computation of
${\rm dim}\,H^0({\bf CP}^n, {\cal O}(1))$ without resorting to sheaf 
cohomology. That is, we compute ${\rm dim}\, {\cal H}$ when $l=1$
and prove that it coincides with the right--hand side of eqn. (\ref{boludo}).
The case $l>1$ will be treated in section \ref{xcomptx}.

Starting with ${\cal C}={\bf CP}^{0}$, {\it i.e.}, a point $p$ as classical phase space, 
the space of quantum rays must also reduce to a point. Then the corresponding Hilbert space 
is ${\cal H}_1={\bf C}$. The only state in ${\cal H}_1$ is the vacuum $|0\rangle_{l=1}$. 
Henceforth, for brevity, we drop the Picard class index from the vacuum. 

Next we pass from ${\cal C}={\bf CP}^0$ to ${\cal C}={\bf CP}^1$. 
Regard $p$, henceforth denoted $p_1$, as the {\it point at infinity}\/ 
with respect to a coordinate chart $({\cal U}_1, z_{(1)})$ on ${\bf CP}^1$ that does not 
contain $p_1$. This chart is biholomorphic to ${\bf C}$ and supports a representation 
of the Heisenberg algebra in terms of creation and annihilation operators $A^{\dagger}(1)$,  
$A(1)$.  On the chart ${\cal U}_1$, the Hilbert space ${\cal H}_2={\bf C}^2$ 
is the linear span of the vacuum  $|0(1)\rangle$ and its excitation $A^{\dagger}(1)|0(1)\rangle$.

On ${\bf CP}^1$ we have the charts $({\cal U}_1, z_{(1)})$ and $({\cal U}_2, z_{(2)})$. 
Point $p_1$ is at infinity with respect to $({\cal U}_1, z_{(1)})$, while it 
belongs to $({\cal U}_2, z_{(2)})$. Similarly, the point at infinity with respect to 
$({\cal U}_2, z_{(2)})$, call it $p_2$, belongs to $({\cal U}_1, z_{(1)})$ but not to 
$({\cal U}_2, z_{(2)})$. On ${\cal U}_2$, the 
fibre is the linear span of $|0(2)\rangle$ and $A^{\dagger}(2)|0(2)\rangle$, 
$A^{\dagger}(2)$ and $\vert 0(2)\rangle$ respectively being the creation operator 
and the vacuum on ${\cal U}_2$. On the common overlap 
${\cal U}_1\cap {\cal U}_2$, the coordinate transformation between $z_{(1)}$ and 
$z_{(2)}$ is biholomorphic. This implies that, on ${\cal U}_1\cap {\cal U}_2$,
the fibre ${\bf C}^2$ can be taken in either of two equivalent ways: either 
as the linear span of $|0(1)\rangle$ and $A^{\dagger}(1)|0(1)\rangle$, or as 
the linear span of $|0(2)\rangle$ and $A^{\dagger}(2)|0(2)\rangle$. 
Indeed, biholomorphicity of the coordinate change between $({\cal U}_1,
z_{(1)})$ and $({\cal U}_2,z_{(2)})$ implies that the vectors
$|0(1)\rangle$ and $A^{\dagger}(1)|0(1)\rangle$ transform bijectively into
the vectors $|0(2)\rangle$ and $A^{\dagger}(2)|0(2)\rangle$, and
viceversa. 

When $n>1$ we proceed by analogy with the case $n=1$. Topologically we have ${\bf CP}^{n}={\bf 
C}^n\cup {\bf CP}^{n-1}$, with ${\bf CP}^{n-1}$ a hyperplane at infinity;
we also need to describe the coordinate charts and their overlaps.
There are coordinate charts $({\cal U}_j, z_{(j)})$, $j=1, \ldots, n+1$ 
and nonempty $f$--fold overlaps $\cap_{j=1}^f {\cal U}_j$ for $f=2,3,\ldots, n+1$. 
Each chart $({\cal U}_j, z_{(j)})$ is biholomorphic with ${\bf C}^n$ and has
a ${\bf CP}^{n-1}$--hyperplane at infinity; the latter is charted by the 
remaining charts $({\cal U}_k, z_{(k)})$, $k\neq j$.
Over $({\cal U}_j, z_{(j)})$ the Hilbert--space bundle ${\cal QH}_{n+1}$ has a fibre 
${\cal H}_{n+1}={\bf C}^{n+1}$ spanned by 
\begin{equation}
|0(j)\rangle,\qquad  A_i^{\dagger}(j) |0(j)\rangle, \qquad i=1,2,\ldots, n. 
\label{pann}
\end{equation}
On every nonempty $f$--fold overlap $\cap_{j=1}^f {\cal U}_j$, the fibre ${\bf C}^{n+1}$ can 
be taken in $f$ different, but equivalent ways, as the linear span of 
$|0(j)\rangle$ and $A_i^{\dagger}(j) |0(j)\rangle$,  $i=1,2,\ldots, n$, for 
every choice of $j=1,\ldots, f$. This is proved by analyticity arguments
analogous to those above, but let us spell out the details in the simple case when $f=2$
(the case $f>2$ involves no novelty with respect to $f=2$).
Assume that ${\cal U}_{j_1}\cap{\cal U}_{j_2}$ is nonempty for some indices $j_1$,
$j_2$. On the chart $({\cal U}_{j_1},z_{(j_1)})$, the fibre ${\bf C}^{n+1}$
is the linear span of the vectors $\vert 0(j_1)\rangle$, $A^{\dagger}_{i_1}(j_1)\vert
0(j_1)\rangle$, for $i_1=1,2,\ldots, n$. Similarly, on  $({\cal U}_{j_2},z_{(j_2)})$, 
the fibre ${\bf C}^{n+1}$ is the linear span of $\vert 0(j_2)\rangle$,
$A^{\dagger}_{i_2}(j_2)\vert 0(j_2)\rangle$, for $i_2=1,2,\ldots, n$. 
The coordinate transformation between $({\cal U}_{j_1},
z_{(j_1)})$ and $({\cal U}_{j_2}, z_{(j_2)})$ is biholomorphic. This implies
that, on the overlap ${\cal U}_{j_1}\cap{\cal U}_{j_2}$, the vectors $\vert 0(j_1)\rangle$, 
$A^{\dagger}_{i_1}(j_1)\vert 0(j_1)\rangle$, $i_1=1,2,\ldots, n$,
transform bijectively into the vectors $\vert 0(j_2)\rangle$,
$A^{\dagger}_{i_2}(j_2)\vert 0(j_2)\rangle$, $i_2=1,2,\ldots, n$, and
viceversa.

A complete description of this bundle requires the specification of the 
transition functions. This will be done in section \ref{mmrrbb}, where
transition functions will be identified with jacobian matrices (for the
coordinate transformations on ${\bf CP}^n$), plus a phase factor. 
Two properties will follow from this fact. The first one is
the cocycle condition, which the transition functions will certainly satisfy. 
The second one is the independence of the bundle with respect to the specific 
coordinates chosen on ${\bf CP}^n$, as long as the coordinates are holomorphic.
In other words, although we have found it convenient to use the particular
holomorphic coordinates $({\cal U}_j, z_{(j)})$ described in section \ref{cipienne}, any other
holomorphic atlas consisting of charts $({\cal W}_j, w_{(j)})$ would have produced 
the same results. In particular, none of the above results depends on the fact that the charts ${\cal U}_j$ 
are biholomorphic to the whole of ${\bf C}^n$. If the new charts ${\cal W}_j$ 
were biholomorphic to open subsets of ${\bf C}^n$ not identical to all of ${\bf
C}^n$, the previous arguments would continue to hold just as well. 
The key property is the biholomorphicity of coordinate transformations on 
overlapping charts, something that is guaranteed by the definition of a
complex manifold. Thus our construction of the ${\cal QH}$--bundle is
functorial, in the sense that it is coordinate--independent.

\subsection{Representations}\label{wrepp}

The $(n+1)$--dimensional Hilbert space of eqn. (\ref{pann}) may be regarded as a kind of 
{\it defining representation}, in the sense of the representation theory 
of $SU(n+1)$ when $n>1$. To make this statement more precise we observe 
that one can replace unitary groups with special unitary groups in eqn. (\ref{facx}).
Comparing our results with those of section \ref{cipienne} we conclude 
that the quantum line bundle ${\cal L}$ now equals $\tau$, 
\begin{equation}
{\cal L}=\tau,
\label{ddttx}
\end{equation}
because $l=1$. This is the smallest value of $l$ that produces a nontrivial 
${\cal H}$, as eqn. (\ref{jjoo}) gives a 1--dimensional Hilbert space when $l=0$. 
So our ${\cal H}$  spans an $(n+1)$--dimensional representation of $SU(n+1)$, that we can 
identify with the defining representation. There is some ambiguity here since the 
dual of the defining representation of $SU(n+1)$ is also $(n+1)$--dimensional. 
This ambiguity is resolved by convening that the latter is generated by the 
holomorphic sections of the {\it dual}\/ quantum line bundle 
\begin{equation}
{\cal L}^*=\tau^{-1}.
\label{ddttxx}
\end{equation}
On the chart ${\cal U}_j$, 
$j=1,\ldots, n+1$, the dual of the defining representation is the linear span of the covectors
\begin{equation}
\langle (j)0|,\qquad  \langle (j)0|A_i(j), \qquad i=1,2,\ldots, n. 
\label{pannx}
\end{equation}
These conclusions must be slightly modified in the limiting case when $n=1$, since all 
$SU(2)$ representations are selfdual. This point will be explained in section \ref{mmrrbb}.

Taking higher representations is equivalent to considering the principal
$SU(n+1)$--bundle (associated with the vector ${\bf C}^{n+1}$--bundle) in a 
representation higher than the defining one. We will see next
that this corresponds to having $l>1$ in our choice of the line bundle $\tau^l$.

\subsection{Computation of ${\rm dim}\,H^0({\bf CP}^n, {\cal O}(l))$}\label{xcomptx}

We extend now our quantum--mechanical computation of ${\rm dim}\,H^0({\bf CP}^n, 
{\cal O}(l))$ to the case $l>1$. As in section \ref{xcompt}, we do not resort to 
sheaf cohomology. The values $l=0,1$ respectively correspond to the trivial and the 
defining representation of $SU(n+1)$. The restriction to nonnegative $l$ follows 
from our convention of assigning the defining representation to $\tau$ and 
its dual to $\tau^{-1}$. 
Higher values $l>1$ correspond to higher representations and can be accounted for as follows. 
Let us rewrite eqn. (\ref{facx}) as  
\begin{equation}
{\bf CP}^{n+l}=SU(n+l+1)/\left(SU(n+l)\times U(1)\right),
\label{xxaa}
\end{equation}
where now $SU(n+l+1)$ and $SU(n+l)$ act on ${\bf C}^{n+l+1}$. Now
$SU(n+l)$ admits $\left({n+l}\atop n\right)$--dimensional representations 
(Young tableaux with a single column of $n$ boxes) that, by restriction, 
are also representations of $SU(n+1)$. Letting $l>1$ vary for fixed $n$,
this reproduces the dimension of eqn. (\ref{jjoo}).

By itself, the existence of $SU(n+1)$ representations with the dimension
of eqn. (\ref{jjoo}) does not prove that, picking $l>1$, the corresponding 
quantum states lie in those $\left({n+l}\atop n\right)$--dimensional representations.
We have to prove that no other value of the dimension fits the given data.
In order to prove it the idea is, roughly speaking, that a value of $l>1$ on 
${\bf CP}^n$ can be traded for $l'=1$ on ${\bf CP}^{n+l}$. That is, an $SU(n+1)$ 
representation higher than the defining one can be traded for the defining 
representation of $SU(n+l+1)$. In this way the ${\cal QH}$--bundle on ${\bf 
CP}^n$ with the Picard class $l'=l$ equals the ${\cal QH}$--bundle on ${\bf 
CP}^{n+l}$ with the Picard class $l'=1$.
On the latter we have $n+l$ excited states ({\it i.e.}, other than the 
vacuum), one for each complex dimension of ${\bf CP}^{n+l}$. We can sort 
them into unordered sets of $n$, which is the number of excited states on 
${\bf CP}^n$, in $\left({n+l}\atop n\right)$ different ways. This selects
a specific dimension for the $SU(n+1)$ representations and rules out the rest.
More precisely, it is only when $n>1$ that some representations are ruled 
out. When $n=1$, {\it i.e.} for $SU(2)$, all representations are allowed, 
since their dimension is $l+1=\left({1+l}\atop 1\right)$. However already for 
$SU(3)$ some representations are thrown out. The number $\left({2+l}\atop 2\right)$
matches the dimension $d(p,q)=(p+1)(q+1)(p+q+2)/2$ of the $(p,q)$ 
irreducible representation if $p=0$ and $l=q$ or $q=0$ and $l=p$,
but arbitrary values of $(p,q)$ are in general not allowed.

To complete our reasoning we have to prove that the quantum line bundle 
${\cal L}=\tau$ on ${\bf CP}^{n+l}$ descends to ${\bf CP}^{n}$ as the $l$--th power 
$\tau^l$. For this we resort to the natural embedding of ${\bf CP}^{n}$ 
into ${\bf CP}^{n+l}$.
Let $({\cal U}_{1}, z_{(1)})$, $\ldots$, $({\cal U}_{n+1}, z_{(n+1)})$ be the 
coordinate charts on ${\bf CP}^{n}$ described in section \ref{cipienne},
and let $(\tilde{\cal U}_{1}, \tilde z_{(1)})$, $\ldots$, $(\tilde{\cal U}_{n+1}, 
\tilde z_{(n+1)})$, $(\tilde{\cal U}_{n+2}, \tilde z_{(n+2)})$, $\ldots$, 
$(\tilde{\cal U}_{n+l+1}, \tilde z_{(n+l+1)})$ 
be charts on ${\bf CP}^{n+l}$ relative to this embedding. This means that the first 
$n+1$ charts on ${\bf CP}^{n+l}$, duly restricted, are also charts on ${\bf CP}^{n}$; 
in fact every chart on ${\bf CP}^n$ is contained $l$ times within ${\bf CP}^{n+l}$. 
Let $t_{jk}(\tau)$, with $j,k=1,\ldots, n+l+1$, be the transition function for $\tau$ 
on the overlap $\tilde{\cal U}_{j}\cap \tilde{\cal U}_{k}$ of ${\bf CP}^{n+l}$. 
In passing from $\tilde{\cal U}_{j}$ to $\tilde{\cal U}_{k}$,
points on the fibre are acted on by $t_{jk}(\tau)$. Due to our choice of embedding, 
the overlap $\tilde{\cal U}_{j}\cap \tilde{\cal U}_{k}$ on ${\bf CP}^{n+l}$ contains 
$l$ copies of the overlap ${\cal U}_{j}\cap {\cal U}_{k}$ on ${\bf CP}^{n}$. 
Thus points on the fibre over ${\bf CP}^n$ are acted on by $(t_{jk}(\tau))^l$, 
where now $j,k$ are restricted to $1,\ldots, n+1$. This means that the line 
bundle on ${\bf CP}^n$ is $\tau^l$ as stated, and the vacuum $|0\rangle_{l'=l}$ 
on ${\bf CP}^n$ equals the vacuum $|0\rangle_{l'=1}$ on ${\bf CP}^{n+l}$. Hence 
there are on ${\bf CP}^n$ as many inequivalent vacua as there are elements in ${\bf 
Z}={\rm Pic}\,({\bf CP}^n)$ (remember that sign reversal $l\rightarrow -l$ 
within ${\rm Pic}\,({\bf CP}^n)$ is the operation of taking the dual 
representation, {\it i.e.}, $\tau\rightarrow \tau^{-1}$).

\subsection{Transition functions}\label{mmrrbb}

At each point $p\in {\bf CP}^n$ there is an isomorphism between the holomorphic cotangent 
space $T_p^*({\bf CP}^n)$ and a complex $n$--dimensional subspace of 
${\cal H}={\bf C}^{n+1}= {\bf C}^n\oplus{\bf C}$, where ${\bf C}^n$ is cotangent to 
${\bf CP}^n$ and ${\bf C}$ is normal to it. As $p$ varies over ${\bf CP}^n$ we have 
the following holomorphic bundles: the quantum Hilbert--space bundle 
${\cal QH}$ (with fibre ${\bf C}^{n+1}$), the cotangent bundle $T^*({\bf CP}^n)$ 
(with fibre ${\bf C}^n$), and the normal bundle $N({\bf CP}^n)$ 
(with fibre ${\bf C}$). Modulo a choice of representation for $T^*({\bf 
CP}^n)$, which will be done below, next we prove that
\begin{equation}
{\cal QH}({\bf CP}^n)=T^*({\bf CP}^n) \oplus N({\bf CP}^n).
\label{adxjqq}
\end{equation}
Eqn. (\ref{adxjqq}) follows from the fact that, in the dual (\ref{pannx}) of the defining 
representation, the operators $A_i(j)$ act as $\partial/\partial z^i_{(j)}$, {\it i.e.}, 
as tangent vectors. Correspondingly, in the defining representation (\ref{pann}), 
their adjoints $A_i^{\dagger}(j)$ in ${\cal H}$ act as multiplication by 
$z^i_{(j)}$. 
Since adjoints in ${\cal H}$ transform as duals on tangent space, the $A_i^{\dagger}(j)$ 
transform as differentials ${\rm d}z^i_{(j)}$, or cotangent vectors. In what follows we will 
identify the cotangent and the tangent bundles, so we can write
\begin{equation}
{\cal QH}({\bf CP}^n)=T({\bf CP}^n) \oplus N({\bf CP}^n),
\label{adxj}
\end{equation}
where $T({\bf CP}^n)$ and $N({\bf CP}^n)$ are subbundles of ${\cal QH}({\bf CP}^n)$.
It follows that tangent vectors to ${\bf CP}^n$ are quantum states 
in (the defining representation of) Hilbert space. In eqn. (\ref{pann}) we 
have given a basis for these states in terms of creation operators acting 
on the vacuum $|0\rangle$. The latter can be regarded as the basis vector 
for the fibre ${\bf C}$ of the line bundle $N({\bf CP}^n)$.

As a holomorphic line bundle, $N({\bf CP}^n)$ is isomorphic to $\tau^l$ for some 
$l\in {\rm Pic}\,({\bf CP}^n)$ $={\bf Z}$. Now the bundle $T({\bf CP}^n)
\oplus N({\bf CP}^n)$ has $SU(n+1)$ as its structure group, which we consider 
in the representation $\rho_l$ corresponding to the Picard class $l\in {\bf Z}$:
\begin{equation}
{\cal QH}_l({\bf CP}^n)=\rho_l(T({\bf CP}^n)) \oplus \tau^l, \qquad l\in {\bf Z}.
\label{adxx}
\end{equation} 
The importance of eqn. (\ref{adxx}) is that it classifies ${\cal QH}$--bundles 
over ${\bf CP}^n$: holomorphic equivalence classes of such bundles are in 1--to--1 
correspondence with the elements of ${\bf Z}={\rm Pic}\,({\bf CP}^n)$. The 
class $l=1$ corresponds to the defining representation of $SU(n+1)$,
\begin{equation}
{\cal QH}_{l=1}({\bf CP}^n)=T({\bf CP}^n)\oplus\tau,
\label{pemex}
\end{equation}
and $l=-1$ to its dual. 
The quantum Hilbert--space bundle over ${\bf CP}^n$ is generally nontrivial, 
although particular values of $l$ may render the direct sum (\ref{adxx}) trivial.
The separate summands $T({\bf CP}^n)$ and $N({\bf CP}^n)$ are both 
nontrivial bundles. 
Nontriviality of $N({\bf CP}^n)$ means that, when $l\neq 0$, the state 
$|0\rangle$ transforms nontrivially (albeit as multiplication by a phase factor)
between different local trivialisations of the bundle. When $l=0$ the vacuum 
transforms trivially.

The preceding discussion also answers the question posed in section \ref{xcompt}: 
what are the transition functions $t({\cal QH}_l)$ for  ${\cal QH}_l$? According to eqn. 
(\ref{adxx}), they decompose as a direct sum of two transition functions, 
one for $\rho_l(T({\bf CP}^n))$, another one for $\tau^l$:
\begin{equation}
t({\cal QH}_l({\bf CP}^n))=t(\rho_l(T{\bf CP}^n))\oplus t({\tau^l}).
\label{decc}
\end{equation}
If the transition functions for $\tau$ are $t(\tau)$, those for 
$\tau^l$ are $(t(\tau))^l$. On the other hand, the transition functions 
$t(\rho_l(T{\bf CP}^n))$ are the jacobian matrices (in representation 
$\rho_l$) corresponding to coordinate changes on ${\bf CP}^n$. 
Then all the  ${\cal QH}_l({\bf CP}^n)$--bundles of eqn. (\ref{adxx}) are nonflat 
because the tangent bundle $T({\bf CP}^n)$ itself is nonflat.

Knowing the transition functions $t({\cal QH}_l({\bf CP}^n))$ we can also answer 
the question posed in section \ref{wrepp} concerning the selfduality of the 
$SU(2)$ representations. It suffices to consider the defining 
representation. The latter is 2--dimensional. By eqn. (\ref{decc}), the 
corresponding transition functions, which are $2\times 2$ complex matrices, 
split block--diagonally into $1\times 1$ blocks, with zero off--diagonal entries.
Hence these matrices are symmetric, {\it i.e.}, invariant under transposition, 
which is the operation involved in passing from a representation to its dual. 
No complex conjugation is involved, since $z\mapsto \bar z$ would involve 
creation and annihilation operators with respect to the antiholomorphic coordinate 
$\bar z$. The notations $A$, $A^{\dagger}$ indicate that, 
if the latter acts as multiplication by a holomorphic coordinate $z$,  
the former acts by differentiation with respect to the {\it same}\/ holomorphic 
coordinate $z$.

\section{Tangent vectors as quantum states}\label{ttvvqqss}

We have seen in section \ref{mmrrbb} that (co)tangent vectors to ${\bf CP}^n$ 
are quantum states. The converse is not true, as exemplified by the vacuum. 
Let us generalise and replace ${\bf CP}^n$ with an arbitrary classical phase space ${\cal C}$. 
We would like to write, as in eqn. (\ref{adxj}), 
\begin{equation}
{\cal QH}({\cal  C})=T({\cal  C}) \oplus N({\cal C}),
\label{adccxj}
\end{equation}
where $N({\cal C})$ is a holomorphic line bundle on ${\cal C}$, whose fibre 
is generated by the vacuum state, and $T({\cal C})$ is the 
holomorphic tangent bundle. Does eqn. (\ref{adccxj}) hold in general?

The answer is trivially affirmative when ${\cal C}$ is an analytic submanifold 
of ${\bf CP}^n$. Such is the case, {\it e.g.}, of the embedding of ${\bf 
CP}^n$ within ${\bf CP}^{n+l}$ considered in section \ref{xcomptx};
Grassmann manifolds provide another example \cite{KN}.
The answer is also affirmative provided that ${\cal C}$ is a 
complex $n$--dimensional, compact, symplectic manifold, whose complex and symplectic 
structures are compatible. Notice that ${\cal C}$ is not required to be K\"ahler;
examples of Hermitian but non--K\"ahler spaces are Hopf manifolds \cite{KN}.
Let $\omega$ denote the symplectic form. Then $\int_{\cal C}\omega^n < \infty$ 
thanks to compactness; this ensures that 
${\rm dim}\, {\cal H}<\infty$. Assuming that the vacuum is nondegenerate, 
as was the case with ${\bf CP}^n$, we can adopt a normalisation similar to that 
of eqn. (\ref{boludo}),
\begin{equation}
\int_{\cal C}\omega^n = n+1,
\label{chepelotudo}
\end{equation}
Let us cover ${\cal C}$ with a {\it finite}\/ set of holomorphic coordinate charts 
$({\cal W}_k, w_{(k)})$, $k=1,\ldots, r$; the existence of such an atlas follows 
from the compactness of ${\cal C}$. We can pick an atlas such that $r$ is 
minimal; compactness implies that $r\geq 2$. 

The construction of the ${\cal QH}({\cal  C})$--bundle proceeds along 
the same lines of section \ref{xcompt}. The chart ${\cal W}_k$ is 
biholomorphic to an open subset of ${\bf C}^n$. The $n$ 
components of the holomorphic coordinates $w^j_{(k)}$, $j=1,\ldots, n$ 
give rise to creation and annihilation operators 
$A_j(k), A_m^{\dagger}(k)$, $j,m=1,\ldots, n$, for every fixed value of $k=1,\ldots, r$. 
The vacuum $|0(k)\rangle_l$ corresponding to $l\in {\rm Pic}\,({\cal C})$, 
plus the $n$ states $A_m^{\dagger}(k)|0(k)\rangle_l$, span 
the fibre ${\bf C}^{n+1}$ of the Hilbert--space bundle over the chart 
${\cal W}_k$.  On overlaps ${\cal W}_j\cap {\cal W}_k$, the fibre can be taken 
in either of two equivalent ways. ${\bf C}^{n+1}$ is either the linear span of $|0(j)\rangle_l$ 
plus the $n$ states $A_m^{\dagger}(j)|0(j)\rangle_l$, or the linear span of $|0(k)\rangle_l$ 
plus the $n$ states $A_m^{\dagger}(k)|0(k)\rangle_l$. Indeed, since the coordinate transformation 
between ${\cal W}_j$ and ${\cal W}_k$ is biholomorphic on ${\cal W}_j\cap{\cal W}_k$, 
the states $|0(j)\rangle_l$, $A_m^{\dagger}(j)|0(j)\rangle_l$ transform bijectively into the states
$|0(k)\rangle_l$, $A_m^{\dagger}(k)|0(k)\rangle_l$. 

Analyticity is central to the above construction of the ${\cal QH}({\cal  C})$--bundle. 
On the contrary, whether the charts ${\cal W}_k$ are biholomorphic to the
whole of ${\bf C}^n$, or only to an open subset strictly contained within ${\bf C}^n$, 
is immaterial. It suffices that ${\cal C}$ be a complex manifold.
Then coordinate transformations are biholomorphic on 
overlapping coordinate charts, and it follows that the above construction 
of the ${\cal QH}({\cal  C})$--bundle is functorial in the sense of
coordinate--independence. Such was already the case with ${\bf CP}^n$. 
Additional parts of section \ref{mmrrbb} concerning ${\bf CP}^n$ 
carry over to ${\cal C}$. Choosing $l\in {\rm Pic}\,({\cal C})$ we determine a holomorphic line bundle 
$N_l({\cal C})$ as in eqn. (\ref{adccxj}), and the latter holds (with a subindex $l$ 
on the left--hand side) under the assumptions made above. By eqn. (\ref{adccxj}) we can 
write for the transition functions 
\begin{equation}
t({\cal QH}_l({\cal  C}))=t(T({\cal C})) \oplus t(N_l({\cal C})),
\label{venxx}
\end{equation}
as we did in eqn. (\ref{decc}). Transition functions for $T({\cal C})$ are jacobian 
matrices, and tangent vectors are quantum states. Holomorphic line bundles such as 
$N_l({\cal C})$ are classified by the Picard group ${\rm Pic}\,({\cal C})$, although 
the latter need not be ${\bf Z}$. Now $T({\cal  C})$ may or may not be trivial. 
If both $T({\cal  C})$ and $N_l({\cal C})$ are trivial, then the full quantum 
Hilbert--space bundle is trivial. A nontrivial ${\cal QH}_l({\cal  
C})$--bundle arises if $T{\cal C}$ 
is nontrivial and this nontriviality cannot be compensated by a nontrivial 
$N_l({\cal C})$, or viceversa. On the other hand ${\cal QH}_l({\cal  C})$ is flat if, 
and only if, both $T({\cal C})$ and $N_l({\cal C})$ are flat.

All these analogies with ${\bf CP}^n$ notwithstanding, it is worth stressing that the construction 
of the ${\cal QH}({\cal  C})$--bundle by no means relies on them. Rather, 
the existence of the ${\cal QH}({\cal  C})$--bundle
is a general result that holds for all complex manifolds 
${\cal C}$ satisfying the requirements stated at the beginning of this section. 
In fact there may also be differences with respect to ${\bf CP}^n$. 
One would like to identify $N_l({\cal C})$ (for some class $l\in {\rm 
Pic}({\cal C})$) with the quantum line bundle ${\cal L}$, but ${\cal C}$ 
need not be quantisable and/or $N_l({\cal C})$ need not possess holomorphic sections. 
Another potential difference is the possible degeneracy of the vacuum.
While all vacua on ${\bf CP}^n$ were nondegenerate, this need not be the 
case on a general ${\cal C}$. We will analyse these issues in a 
forthcoming article.

\section{Discussion}\label{dicu}

Quantum mechanics is defined on a Hilbert space of states whose construction
usually assumes a global character on classical phase space. Under {\it 
globality}\/ we understand, as explained in section \ref{intro}, the 
property that all coordinate charts on classical phase space are quantised 
in the same way. A novelty of our approach is the local character of the Hilbert space:
there is one on top of each Darboux coordinate chart on classical phase space.
The patching together of these Hilbert--space fibres on top of each chart may be global 
(trivial bundle) or local (nontrivial bundle). In order to implement 
duality transformations we need a nonflat bundle (hence nontrivial).
Flatness would allow for a canonical identification, by means of parallel 
transport, of the quantum states belonging to different fibres.

Given a classical phase space as a base manifold and a Hilbert space as a fibre, 
the trivial bundle corresponding to these data is unique. On the contrary, 
there may be more than one (equivalence class of) nonflat (and hence 
nontrivial) bundles possessing the given base and fibre. This means that, 
considering nonflat bundles, the choice of a quantum mechanics need not be unique, 
even if the corresponding classical mechanics is kept fixed. The freedom in choosing 
different Hilbert--space bundles is parametrised by the Picard group of classical 
phase space. This group parametrises (equivalence classes of) holomorphic line bundles. 
The corresponding 1--dimensional fibre is spanned by the vacuum state. The remaining 
quantum states are obtained by the action of creation operators on the vacuum chosen. 
The quantum states so obtained can be identified with tangent vectors to classical 
phase space. When the Picard group is trivial, there exists just one Hilbert--space bundle 
(though not necessarily trivial). A nontrivial Picard group means that there 
is more than one equivalence class of Hilbert--space bundles. Any two different choices 
of a Hilbert--space bundle correspond to two different choices of a line 
bundle on which the vacuum state lies.
The previous conclusions are valid on an arbitrary complex, compact classical phase space 
whose complex structure is kept fixed and is compatible with the symplectic 
structure, and assuming nondegeneracy of the vacuum.

In the presence of a nontrivial Picard group, each choice of a line bundle carries 
with it the choice of a representation for the unitary structure group of the 
Hilbert--space bundle. This may lead to the {\it wrong}\/ conclusion that 
duality transformations are just different choices of a representation for 
the unitary group of Hilbert space. A choice of representation 
is {\it not}\/ a duality transformation. The choice of a 
representation for the unitary group is subordinate to the choice of a class 
in the Picard group. Picking a class in the latter, one determines a representation 
for the former. In other words, in eqn. (\ref{adxx}), one does not vary the 
representation $\rho_l$ independently of the Picard class $l$.

A duality thus arises as the possibility of having two or more, apparently different, 
quantum--mechanical descriptions of the same physics. Mathematically, a 
duality arises as a nonflat, quantum Hilbert--space bundle over 
classical phase space. This notion implies that the concept of a quantum is not absolute, 
but relative to the quantum theory used to measure it \cite{VAFA}. That is, duality 
expresses the relativity of the concept of a quantum. In particular, {\it 
classical}\/ and {\it quantum}\/, for long known to be deeply related \cite{PERELOMOV} 
are not necessarily always the same for all observers on phase space.

{\bf Acknowledgements}

It is a great pleasure to thank J. de Azc\'arraga for encouragement and support. Technical discussions with U. Bruzzo and M. Schlichenmaier are gratefully acknowledged. The author thanks S. Theisen and Max-Planck-Institut f\"ur Gravitationsphysik, Albert-Einstein-Institut, for hospitality. 
This work has been partially supported by research grant BFM2002--03681 from Ministerio de Ciencia y Tecnolog\'{\i}a, by EU FEDER funds, by Fundaci\'on Marina Bueno and by Deutsche Forschungsgemeinschaft.

\end{document}